# Dynamical Evolution of Anisotropic Response in Black Phosphorus under Ultrafast Photoexcitation

Shaofeng Ge[1,2],Chaokai Li[1,2], Zhimin Zhang[1,2],Chenlong Zhang[1,2], Yudao Zhang[1,2], Jun Qiu[1,2], Qinsheng Wang[1,2], Junku Liu[1,3], Shuang Jia[1,2], Ji Feng[1,2,*], Dong Sun[1,2,*]

[1]International Center for Quantum Materials, School of Physics, Peking University, Beijing 100871, P. R. China
[2]Collaborative Innovation Center of Quantum Matter, Beijing 100871,P. R. China
[3]Qian Xuesen Laboratory of Space Technology, China Academy of Space Technology, Beijing 100094, P. R. China
[*]Email: jfeng11@pku.edu.cn (J. F.), sundong@pku.edu.cn (D.S.);

Abstract
Black phosphorus has recently emerged as a promising material for high performance electronic and optoelectronic device for its high mobility, tunable mid-infrared bandgap and anisotropic electronic properties. Dynamical evolution of photo excited carriers and its induced change of transient electronic properties are critical for materials' high field performance, but remains to be explored for black phosphorus. In this work, we perform angle-resolved transient reflection spectroscopy to study the dynamical evolution of anisotropic properties of black phosphorus under photo excitation. We find that the anisotropy of reflectivity is enhanced in the pump induced quasi-equilibrium state, suggesting an extraordinary enhancement of the anisotropy in dynamical conductivity in hot carrier dominated regime. These results raise enormous possibilities of creating high field, angle sensitive electronic, optoelectronic and remote sensing devices exploiting the dynamical electronic anisotropic with black phosphorus.

As a bridge 2D material between graphene and transition metal dichalcogenide in terms of mobility and bandgap opening[1,2], BP processes relatively high mobility (up to 50000 cm /(V s))[3-5] and thickness dependent bandgap ranging from near-infrared (single layer) to mid-infrared (bulk)[6-8]. All these fascinating properties make BP highly promising for applications such as high speed electronic device with high on-off ratio[9-13] and mid-infrared optoelectronics device with low noise[2,14-19]. In addition to these attractive properties, BP also exhibits unusual anisotropic in-plane conductivities resulting from orthorhombic lattice symmetry, which allows for a new degree of freedom in designing conceptually novel optoelectronic, electronic and optomechanical devices that exploit electronic anisotropy[2,5,6,15,19-22]. Indeed, this electronic anisotropy affords novel device paradigm impossible in graphene and transition metal dichalcogenide. To facilitate device design based on this novel paradigm, the anisotropic steady-state response of BP has been recently investigated using many angle-resolved characterization techniques, such as transmission spectroscopy[2,7,19], Raman spectroscopy[2,6], photoconductivity measurement[2] and photoluminescence measurement[6]. While steady-state transport is governed by the carriers near the Fermi level, transport in high speed devices is determined by the dynamic conductivity of hot carriers, i.e., electrons whose temperature is elevated above the lattice temperature due to the presence of high fields and/or dynamic fields in the material[23,24]. Despite its importance for high performance device, the dynamical anisotropic response of BP under high field has yet to be investigated.

In this work, we performed angle-resolved pump-probe reflection spectroscopy to study the dynamical response of BP in the presence of photo excitation of hot carriers. The optical excitation with ultrafast pump pulse mimics the working environment of optoelectronic and electronic device operating in the presence of high-field and/or dynamic field. In this regime, carriers are accelerated to elevated energy state, and thus the excited carriers in quasi-equilibrium distributions dominate the device performances. For example, in a typical high speed Si based device with velocity saturation of $2x10^7$ cm/s, the carrier energy is elevated by up to 0.1 eV.[25,26] Therefore it's interesting to examine the dynamical response of BP under photoexcitation, especially the evolution of electronic anisotropy, through polarization-resolved transient reflection measurements. The most remarkable finding inferred from the dynamics of differential reflectance of BP is that not only the anisotropic response of BP is preserved but also the degree of anisotropy is enhanced in the quasi-equilibrium excited states of a high field device. By systematically varying the combination of pump and probe polarizations, it is revealed that the enhancement of electronic anisotropy in photoexcited quasi-equilibrium states is very robust and largely independent of the polarization of the pump photoexcitation. The anisotropic stretch of the conductivity persists until the hot carriers relax via isotropic cooling and recombination.

## Results

### Characterization of BP thin-film and Experimental Scheme

Bulk BP is a layered material with orthorhombic crystal structure[27] as shown in Fig. 1a. In contrast to graphene, each layer in BP has puckered lattice structure which results in its unique angle-dependent in-plane conductivities[2] due to reduced symmetry. Optical micrograph of a typical thin BP flake is shown in Fig. 1c. The majority of later measurements presented in this work are performed on a sample with thickness of 25 nm determined by atomic force microscopy (Fig. 1d), from which we can deduce the thickness to be about 47 layers using a layer spacing of 0.53 nm[7]. The as-synthesized BP is found to be p-doped with doping density of 4.2 x $10^{18}/cm^3$ by Hall measurement, equivalent to 1.05 x $10^{13}/cm^2$ for a 25-nm thick sample (2 x $10^{11}/cm^2$ for each BP layer). Polarization resolved transient reflection measurement is performed on samples with thickness ranging from 10-30 nm in this work and the results are qualitatively similar.

A schematic diagram of the experiment is shown in Fig. 1a. A 1.55-eV linear polarized pump laser are used to induce photo carriers in BP. Typical pump fluence is $3 \times 10^{15}$ photons/$cm^2$ which converts to $3 \times 10^{14}/cm^2$ electron-hole pairs at the initial stage using a 0.4%/nm absorption rate at 800 nm[7]. The photoexcited carriers cause quasi-Fermi level shift of 354 meV in conduction band and 225 meV in valence band from a steady state band filling consideration assuming no Fermi level smearing (0 K) in the sample. Due to the photo excitation, the anisotropic optical conductivity of BP is expected to be modified. To measure dynamical evolution of the anisotropic optical conductivity, polarization resolved 647-meV probe photon arriving at varies delay time (t) with respect to pump pulse is used, with which the pump induced reflection change (ΔR) is measured. The probe photon energy is only 340 meV larger than the bandgap[8], which corresponds primarily to the transition involving lowest conduction and highest valence band. The probe polarization is rotatable by a half wave plate, so the anisotropic response of BP can be determined dynamically with a 200 fs time resolution as limited by the convolution of pump-probe pulse.

Figure 1e shows a typical transient reflection spectrum with pump and probe polarization set at 0° and 90° respectively (angle defined in Fig. 1c). The initial $\Delta R/R|_{t=0}$ is negative, which indicates that pump-induced reflection decreases. Subsequently, ΔR/R decays gradually. However, ΔR/R reaches a nonzero constant negative value that persists for well beyond 1 ns after the initial pump. This long-lived background is due to elevation of lattice temperature, and the relaxation of this background is attributed to the cooling of lattice temperature of BP through dissipation of heat to the substrate (see supplementary information). In later analysis, we have subtracted the constant background from ΔR/R.

**Pump polarization and fluence dependence of transient reflection spectra**

Figure 2a shows the pump polarization dependent transient reflection spectra (normalized at $\Delta R/R|_{t=0}$) with probe polarization fixed along 0° and 90° . The transient reflection at time zero ($\Delta R/R|_{t=0}$) shows obvious pump polarization dependence. If we assume $\Delta R/R|_{t=0}$ has linear dependence on photoexcited carrier density, we can simply fit $\Delta R/R|_{t=0}$ as function of pump polarization angle $\alpha_1$ (respective to x axis as marked in Fig. 1c) with polarization dependent absorption coefficient A[7](see supplementary information for derivation):

$$A \approx \frac{4\sqrt{\epsilon_1}(Re(\sigma_{xx})\cos^2(\alpha_1)+\mathrm{Re}(\sigma_{yy})\sin^2(\alpha_1))}{\epsilon_0 c(\sqrt{\epsilon_2}+\sqrt{\epsilon_1})^2} \qquad (1)$$

Here we assume BP layers are sandwiched between two dielectric media, i.e. substrate and air, with dielectric constant of $\varepsilon_1$ and $\varepsilon_2$, where $\sigma_{xx}$ and $\sigma_{yy}$ are nonzero diagonal component of optical conductivity tensor, $\varepsilon_0$ is the free-space permittivity and c is the speed of light. The fitting result is shown in Fig. 2b, we can determine the angle between the x crystal axis of BP and the blue arrow (0° in Fig. 1c) to be -5.3 degree. Similar fitting can be performed on $\Delta R/R|_{t=0}$ with probe polarization fix along 0° , which gives the same angle of x crystal axis of BP. The linear dependence of $\Delta R/R|_{t=0}$ on photoexcited carrier density is verified by pump fluence dependent measurement as shown in Fig. 2d. $\Delta R/R|_{t=0}$ shows clear linear dependence on pump fluence regardless of specific pump probe polarization. This suggests the measured amplitude variance of differential reflectivity as function of pump polarization is purely a manifestation of anisotropic absorption of pump light.

To study the carrier densities dependent relaxation dynamics of ΔR/R, we try to fit relaxation dynamics of ΔR/R thereafter. Exponential decay fittings would not only require at least two exponential terms but also a changing of ratio between different components with pump fluence (supplementary information), this indicates nonlinear carrier density dependent recombination process. In view of this, we fit the entire set of data of the transient spectrum at different pump fluence by a simple model based on rate equation[28]:

$$\frac{dN_n(t)}{dt} = G(t) - AN_n(t) - BN_n^2(t) - CN_n^3(t) \qquad (2)$$

where $N_n(t)$ is the electron density, A ($s^{-1}$), B ($cm^3/s$), and C ($cm^6/s$) are the Schockley-Reed, direct electron-hole and Auger recombination coefficients[29], respectively. The generation rate, G ($cm^{-3}$ $s^{-1}$), is proportional to the absorbed laser excitation q ($W/cm^2$) which is nonzero only during the pump pulse duration. Fig. 2c shows the fit to the complete set of pump fluence dependent measurement with uniform A, B, C parameters, the fitting results are plotted as solid lines. The relative contribution of Schockley-Reed, direct electron-hole and Auger recombination to the carrier density relaxation is plotted in supplementary information, Schockley-Reed and radiative recombination dominates the relaxation process, whereas the Auger process is negligible except during the initial stage after the pump excitation. We note this simple fitting is intended for qualitative evaluation of the relative proportion of lowest order recombination pathways, instead of quantitative determination of the actually dynamics.

**Dynamical probing of pump induced anisotropic response**

Figure 3a shows the probe polarization dependence of transient reflection spectra with pump polarization fixed along 0° and 90°. As the probe polarization rotates from 0° to 90°, the $\Delta R/R|_{t=0}$ stays negative which corresponds to a pump induced reflection decrease in both x and y axis of BP. The negative ΔR/R persists for several picoseconds, after which the value switches to positive for polarization angle close to x-axis. Further experiment shows the evolution of spectra as function of probe polarization is qualitatively the same regardless of pump polarization; Consequently, we confirm that the transient reflection spectra depends weakly on pump polarization solely through anisotropic absorption. The material itself doesn't have any "memory" of pump polarization for the 200-fs experimental time resolution. As the initial negative ΔR/R around time zero is possibly due to highly non-equilibrium state, we normalize the peak negative/positive ΔR/R signal starting from ~4.5 ps to examine the relaxation dynamics at different probe polarizations. As shown in Fig. 3b, the relaxation dynamics after the initial non-equilibrium state, is independent of the probe polarization, which indicates the recombination and relaxation dynamics of photoexcited carriers are isotropic in BP.

The most remarkable observation is the sinusoidal oscillation of the ΔR/R as function of probe polarization at fixed pump-probe delay, as shown in Fig. 3c. This periodic oscillation is robust and persists for the whole decay process (~ 1 ns) except for the initial highly non-equilibrium state. This can be directly explained by the anisotropic response of optical conductivity. The measured ΔR/R can be converted to optical conductivity change Δσ through the following equation assuming $\Delta\sigma \ll \sigma$ which is valid after the initial relaxation of highly excited states: (see supplementary information for deduction and the parameter range that the approximation is valid)

$$\frac{\Delta R}{R} \approx \frac{4\sqrt{\epsilon_1}\Delta Re\sigma}{\epsilon_0 c(\epsilon_2-\epsilon_1)} + \frac{8\sqrt{\epsilon_1\epsilon_2}(\Delta Im\sigma)Im\sigma}{\epsilon_0{}^2c^2(\epsilon_2-\epsilon_1)^2} \approx \frac{4\sqrt{\epsilon_1}(\Delta Re(\sigma_{xx})cos^2\alpha_2+\Delta Re(\sigma_{yy})sin^2\alpha_2)}{\epsilon_0 c(\epsilon_2-\epsilon_1)} + \frac{8\sqrt{\epsilon_1\epsilon_2}\left((\Delta Im(\sigma_{xx})cos^2\alpha_2+\Delta Im(\sigma_{yy})sin^2\alpha_2)\right)Im\sigma}{\epsilon_0{}^2c^2(\epsilon_2-\epsilon_1)^2} \quad (3)$$

The first and the second term of Eq. (3), respectively, account for the contribution of $\Delta\mathrm{Re}(\sigma)$ and $\Delta\mathrm{Im}(\sigma)$ (pump induced change of the real and imaginary parts of optical conductivity) to the transient reflection. In both terms, the transient reflection has sinusoidal dependence on the probe polarization angle $\alpha_2$. Thus, from Eq. (3), to account for the sinusoidal oscillation observed in Fig. 3b, it follows immediately that $\Delta\sigma_{xx} \neq \Delta\sigma_{yy}$. This indicates Δσ is anisotropic in BP although the relaxation dynamics are isotropic.

**Dynamical evolution of optical conductivity at quasi-equilibrium photo excited state**

To interpret the dynamical response of BP under photo excitation, we use a two-band model[7,30] to compute the optical conductivity of BP in quasi-equilibrium states, as a function of the effective temperature and the photo-generated excess electron-hole density. The effective temperatures of electron and hole gases as well as their chemical potentials vary with time, as the density decreases through recombination. The real parts of the conductivity tensor are shown in Fig. 4. The simulation gives Re($\sigma_x$) =4.84 and Re($\sigma_y$)= 0.37 for unexcited state with initially hole doping of 1.03*$10^{13}$/cm$^2$. Comparing with unexcited value, both ΔRe($\sigma_{xx}$) and ΔRe($\sigma_{yy}$) are negative in the parameter range that is plotted. The change of Re ($\sigma$) is primarily due to the Pauli blocking effect: the real part of conductivity, which mainly comes from interband transition, will be reduced due to the increased occupation of conduction and valance band of photo excited carriers. We also observe the Re($\sigma$) generally decreases with increasing photo carrier density, on the other hand, it shows a minimum when temperature increases at fixed density. This is expected from the effect of Pauli blocking when the carrier density is well below the energy scale of the probe transition. What is not taken into account in this model is the possible gap renormalization due to the electron hole interaction[31] , which could enhance the interband transition and thereby even over-compensate the Pauli Blocking.

The dynamical evolution of conductivity in parameter regions that is relevant to high field device can be extracted directly from experimentally measured transient reflection signal, independent of the above model. We note although both ΔRe(σ) and ΔIm(σ) as well as their respective contributions to the transient reflection vary throughout the relaxation processes, the contribution from ΔRe(σ) usually dominates (See Fig. S6 of supplementary information). This is expected from Eq. (3), as the coefficient of the first term: $\frac{4\sqrt{\epsilon_1}}{\epsilon_0 c(\epsilon_2-\epsilon_1)} \sim \frac{0.03}{\sigma_0}$ is far larger than the coefficient of the second term: $\frac{8\sqrt{\epsilon_1\epsilon_2} Im\sigma}{\epsilon_0{}^2 c^2(\epsilon_2-\epsilon_1)^2} \sim \frac{0.001}{\sigma_0} * \frac{Im\sigma}{\sigma_0}$, taking $\sigma_0 = e^2/4\hbar$, $\epsilon_1 = 1, \epsilon_2 = 3.9$, $\epsilon_0 c = \frac{\sigma_0}{0.023} = 43.5\,\sigma_0$, and Im $\sigma_x = -0.52\,\sigma_0$ and Im $\sigma_y = -0.60\ \sigma_0$ for native state of 25 nm thick BP sample (see Methods). Thus, in the limit ΔRe (σ) is comparable with or larger than ΔIm(σ), which is valid over the quasi-equilibrium state region after initial fast carrier cooling and recombination (as marked in Figure S6 of supplementary information), the second term of Eq. (3) can be ignored. This simplification allows us to extract the dynamical evolution of Re (σ) through the simple relationship $\frac{\Delta R}{R} \approx \frac{4\sqrt{\epsilon_1}\Delta Re\sigma}{\epsilon_0 c(\epsilon_2-\epsilon_1)}$ from experimentally measured transient reflection signal, which is shown in Fig. 3d and e. We observed conductivity ellipse is stretched (ΔRe($\sigma_{xx}$) > 0 while ΔRe ($\sigma_{yy}$) < 0) due to photoexcitation of hot carriers, the stretch of conductivity ellipse persists over the subsequent relaxation process and gradually recovers as the photo excited hot carriers relax and recombine.

The stretch of conductivity ellipse implies BP becomes more anisotropic under high field transport, which is equivalent to a quasi-equilibrium state of photo excited carriers after initial relaxation and recombination processes. Here we want to clarify that the deduced conductivity ellipse closed to timezero (0 ps , 0.4 ps and possibly 4.3 ps delay) is dominated by either non-equilibrium state or highly excited quasi-equilibrium state. During this stage, the condition ΔRe(σ) ~ ΔIm(σ) or ΔRe(σ)>ΔIm(σ) is invalid, and the contribution of ΔIm(σ) should not be ignored and may dominate the contribution to transient reflection signal. The dynamical evolution of conductivity is hard to justify from experiment in this region. However, if using typical energy of 0.1 eV as carriers accelerated to velocity saturation in high field device, this energy is equivalent to 1000 K or photo induced doping intensity of 3*$10^{13}$/cm$^2$ if filling from the bottom of conduction band. This falls into the parameters region of the long transient reflection tails after the initial relaxation and recombination of photo carriers which is irrelevant to high speed device application.

**Discussions**

Here we want to emphasize the enhanced anisotropic response is not relevant to pump polarization. Although the absorption of photon through direct interband transition in BP has linear dichroism, the memory of the BP of the excitation polarization is shorter than the ~200 fs time resolution of our experiment. The non-uniform distribution of carrier in k space is erased possibly due to fast carrier-carrier scattering process. The isotropic relaxation along any polarization directions thereafter may also share the same origin as above: the fast carrier-carrier scattering mediates the carrier distribution in k space in sub-200 fs timescale, this fast modification of carrier distribution gives isotropic relaxation/recombination behavior magnified in the transient reflection spectroscopy, although the carrier phonon scattering and carrier recombination are possibly highly anisotropic.

It should be remarked that the computed change in the real part of the conductivity in x-direction is ostensibly negative, although the experimentally inferred $\Delta\mathrm{Re}\,(\sigma_{xx})>0$. There are a few sources that can in principle cause this discrepancy, from the modeling perspective. First, the computed $\Delta\mathrm{Re}\,(\sigma_{xx})$ is negative primarily because of the Pauli Blocking. First, the gap renomarlization due to the many body effect, as mentioned earlier, could compensate or even over compensate the Pauli blocking. Second, there could be multi-band physics that are beyond the scope of this simple two-band model. These, however, make the microscopic origin of the enhancement of electronic anisotropy all the more interesting. Further theoretical analysis is warranted to clarify the dynamical processes of photo-carriers as well as change in the electronic structure in the pump-probe experimental setting.

This experimental work provides interesting device physics toward understanding angle sensitive operation behavior of various BP based optoelectronics device when light illumination is involved, such as a photodetector, remote sensing device and optical modulator or any BP based devices running in high field transport limit. On top of BP's extraordinary properties of high mobility and thickness dependent bandgap suitable for mid-infrared wavelength, a spectral range that was previously difficult to access, the preservation and enhancement of anisotropic response and linear dichroism under excited state makes BP an ideal material, distinguished from conventional materials and approaches for angle sensitive devices, capable of working in high-field applications such as ballistic transistor and high speed polarization sensors where both high mobility and anisotropy are desired. As a new member of 2D family, BP also shares the advantage of easy integration with other 2D materials, facilitating its application for multifunctional integrated electronics and photonics circuits based on multiple 2D materials.

## Methods

**Sample Preparation:** Thin black phosphorus（BP）flakes are obtained by mechanical exfoliation a synthesized bulk BP crystal on 285 nm SiO2 substrate. Single crystals of black phosphorus(BP) are grown by a chemical vapor transport (CVT) method as previously described[32]. Red phosphorus (500mg), AuSn (364mg) and SnI4 (10mg) are sealed in an evacuated quartz ampoule of 12cm long. The end with the charges of the ampoule is placed horizontally at the center of a single zone tube furnace. The ampoule is slowly heated up to 873K within 10h and kept at this temperature for 24h. Then the ampoule was cooled to 773K at a rate of 40K/h. Single crystals of BP crystallized in flake-like form with the sizes larger than 1cm are obtained at the cold end of the ampoule.

**Transient Reflection Spectroscopy:** An infrared optical parametric amplifier (OPA) pumped by a 60 fs, 250 kHz Ti: Sapphire regenerative amplifier (RegA) is usedin the transient reflection measurement. The idler from OPA at 1940 nm (~150fs) is used as probe and the dispersion-compensated residual 800 nm of the OPA is used as the pump. Both pump and probe pulses are linearly polarized and two half-wave plates are used to alter their polarization angle respectively. A 40X reflective objective lens is used to focus the co-propagating pump probe spots onto the sample which is cooled in a liquid-nitrogen-cooled cryostat for temperature control. The reflected probe is collected by the same objective lens and routed through a monochromator followed by an InGaAs photodetector. The detected probe reflection isread by lock-in amplifier referenced to 5.7 kHz mechanically chopped pump. The probe spot size is estimated to be 4 μm with a pump spot size slightly larger. The pump excited carrier density is estimated to be around $1.3 \times 10^{15}/cm^2$ unless in a pump fluence dependent measurement.

**Two-band Model Simulation:** We use the two-band tight binding (TB) model as in Ref.[7,30]:

$$H_j(\boldsymbol{k}) = \begin{bmatrix} E_{cj} + \eta_c k_x^2 + \nu_c k_y^2 & \gamma k_x + \beta k_y^2 \\ \gamma k_x + \beta k_y^2 & E_{vj} - \eta_v k_x^2 - \nu_v k_y^2 \end{bmatrix}$$

where $j$ is the subband index, $\eta_c = \eta_v = \hbar^2/0.4m_0$, $\nu_c = \hbar^2/1.4m_0$, $\nu_v = \hbar^2/2.0m_0$, $\gamma = 4a/\pi$ eV, $\beta = 6.2\, a^2/\pi^2$eV, $m_0$ is the free electron mass, $2a = 4.46$Å is the lattice constant along the x-direction. Because the sample is 25 nm thick, the energy dispersion of the z-direction should be considered. By the envelope function method, $E_{cj} = j^2\pi^2\hbar^2/2m_z^c d^2$, $E_{vj} = -\delta(d) - j^2\pi^2\hbar^2/2m_z^v d^2$, where $m_z^c = 0.2m_0$, $m_z^v = 0.4m_0$, $d = 25$ nm is the sample thickness , and $\delta(d) = 0.3$eV is the band gap.

The conductivity corresponding to the incident light with frequency ω is given by the Kubo's formula. The scattering time is set to the equivalent of 0.01 eV, following Low et al[7,30]. For equilibrium states, the chemical potentials of electrons and holes are equal:$\mu_e = \mu_h$. However, when the photoexcited electrons and holes settle down to a quasi-equilibrium state, their chemical potentials are in general unequal: $\mu_e \neq \mu_h$. $\mu_e$ and $\mu_h$ are determined by the photo-carrier density and their quasi-equilibrium temperature. For all calculations, the BP is hole-doped with a hole density of $10^{13}$ cm$^{-2}$.

**Acknowledgement:**

This project has been supported by the National Basic Research Program of China (973 Grant Nos. 2012CB921300, 2013CB921900, 2014CB920900), the National Natural Science Foundation of China (NSFC Grant Nos. 11274015 and 11174009), the Recruitment Program of Global Experts, Beijing Natural Science Foundation (Grant No. 4142024) and the Specialized Research Fund for the Doctoral Program of Higher Education of China (Grant No.20120001110066).

**Figures and Figure Captions**

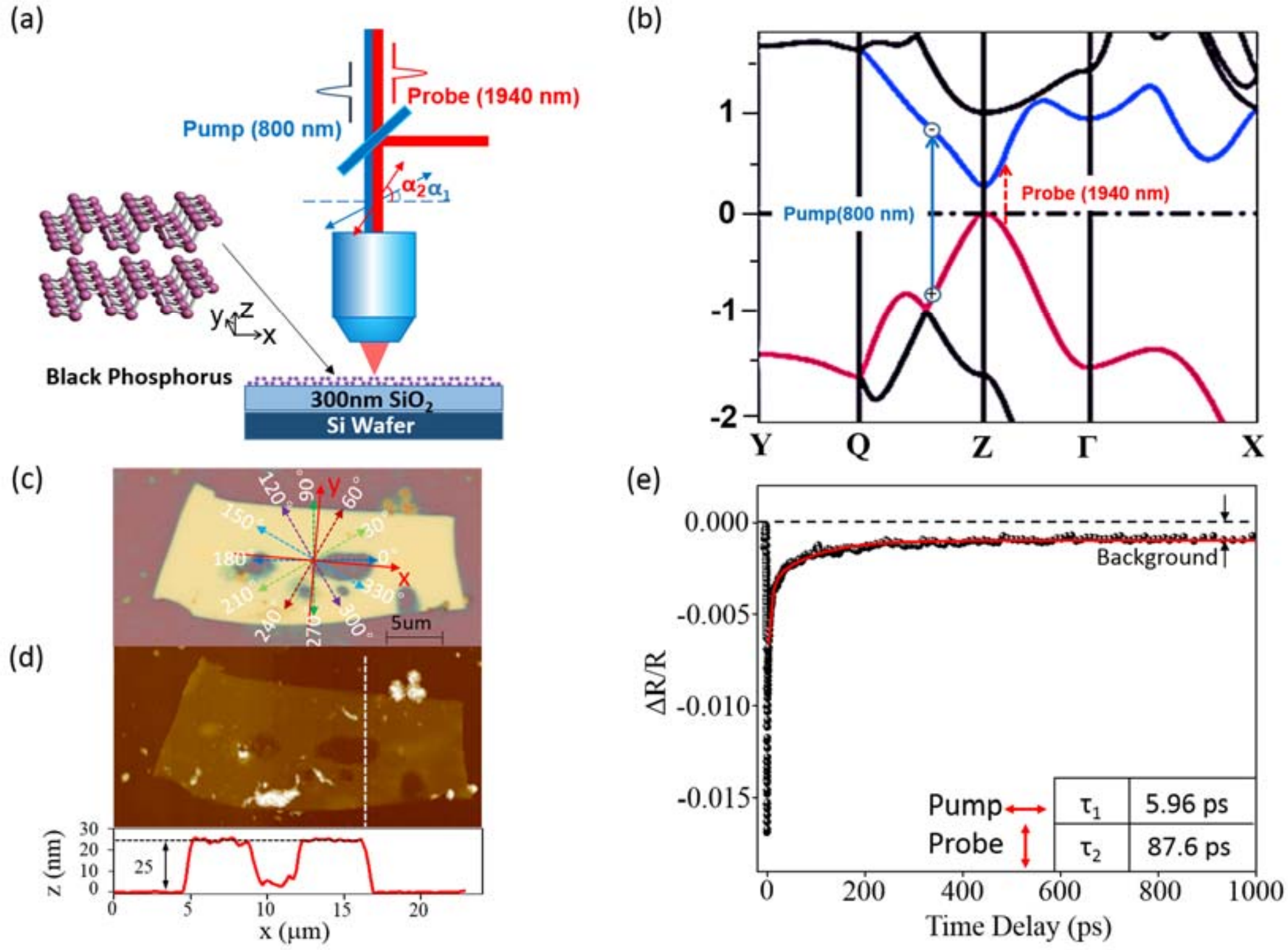

**Figure 1. Experimental scheme and sample characterization** (a) Schematic diagram of polarization resolved transient reflection experiment. (b) Band diagram of black phosphorus and pump (blue) probe (red) photon transition configuration. (c) Optical micrograph (d) Atomic force micrograph (e) Representative transient reflection spectrum with pump polarization along 0˚ (~ x-direction in Fig. c)) and probe polarization at 90˚ (~ y-direction in Fig. c). The data is fitted by bi-exponential decay function with two decay time constants: $\tau_1, \tau_2$.

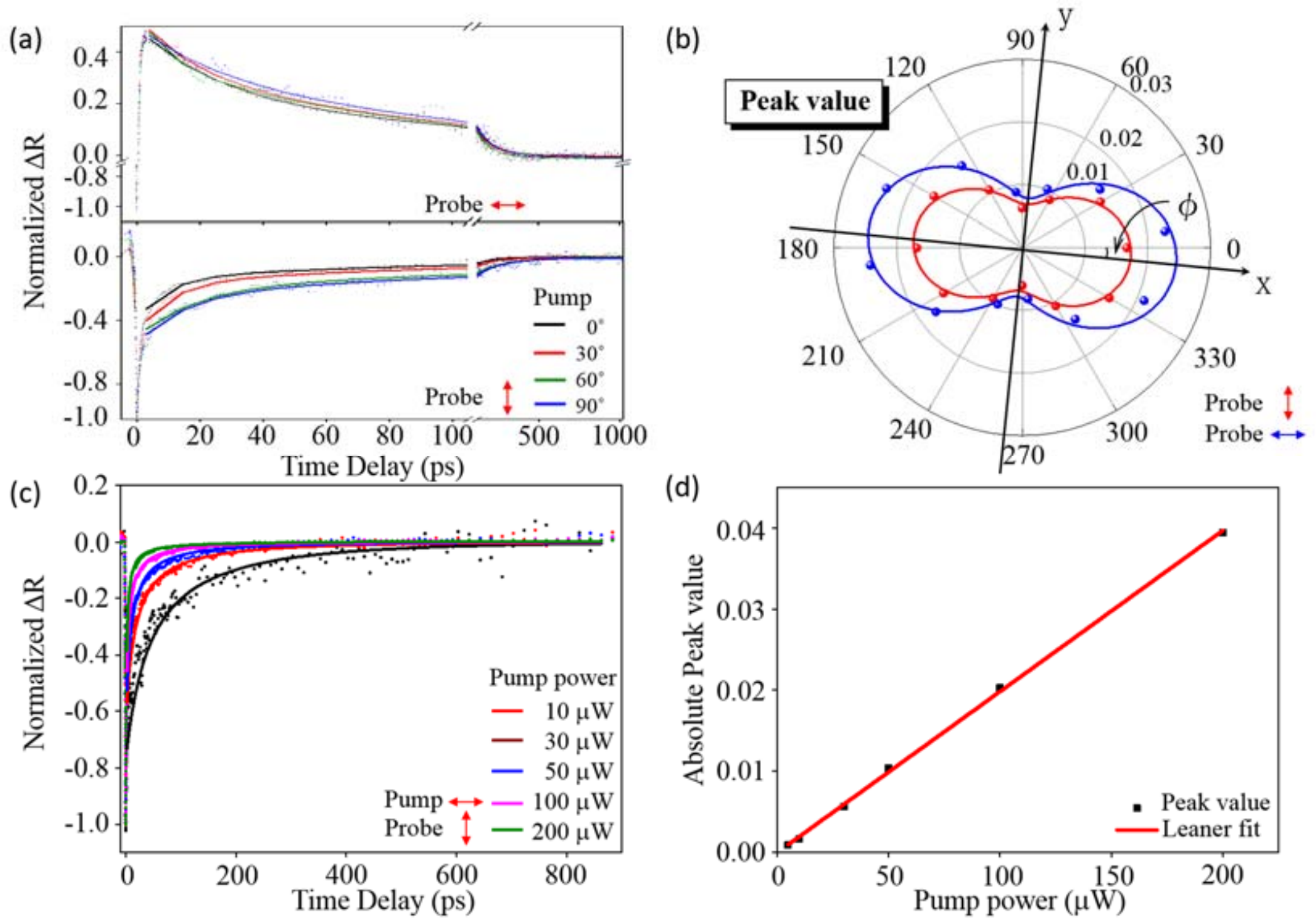

**Figure 2. Pump polarization and fluence dependence of transient reflection spectra** (a) Transient reflection spectrum (normalized) with pump laser polarization at 0° ,30° ,60° ,90° with probe polarization fixed along 0° and 90°. The pump wavelength is 800 nm and the probe wavelength is 1940 nm. The results are fitted by the bi-exponential decay function. To compare the decay dynamics, the results are normalized by their negative peak values. (b) $\Delta R/R|_{t=0}$ as function of pump polarization with probe polarization fixed at 0°(blue dots) and 90°(red dots). The polarization dependence of $\Delta R/R|_{t=0}$ can be fit by the function $A_1 cos^2(\theta+\varphi) + A_2 sin^2(\theta+\varphi)$ where φ is the angle between the crystal x direction and the 0° as marked in Figure 1c. The fitting gives the φ= -5.2° for probe polarization at 0° and -5.3° for probe polarization at 90°. (c) Normalized ΔR with different pump excitation power (pump polarization of 0° and probe polarization of 90°). (d) $\Delta R/R|_{t=0}$ as function of pump power. The red line is the leaner fit.

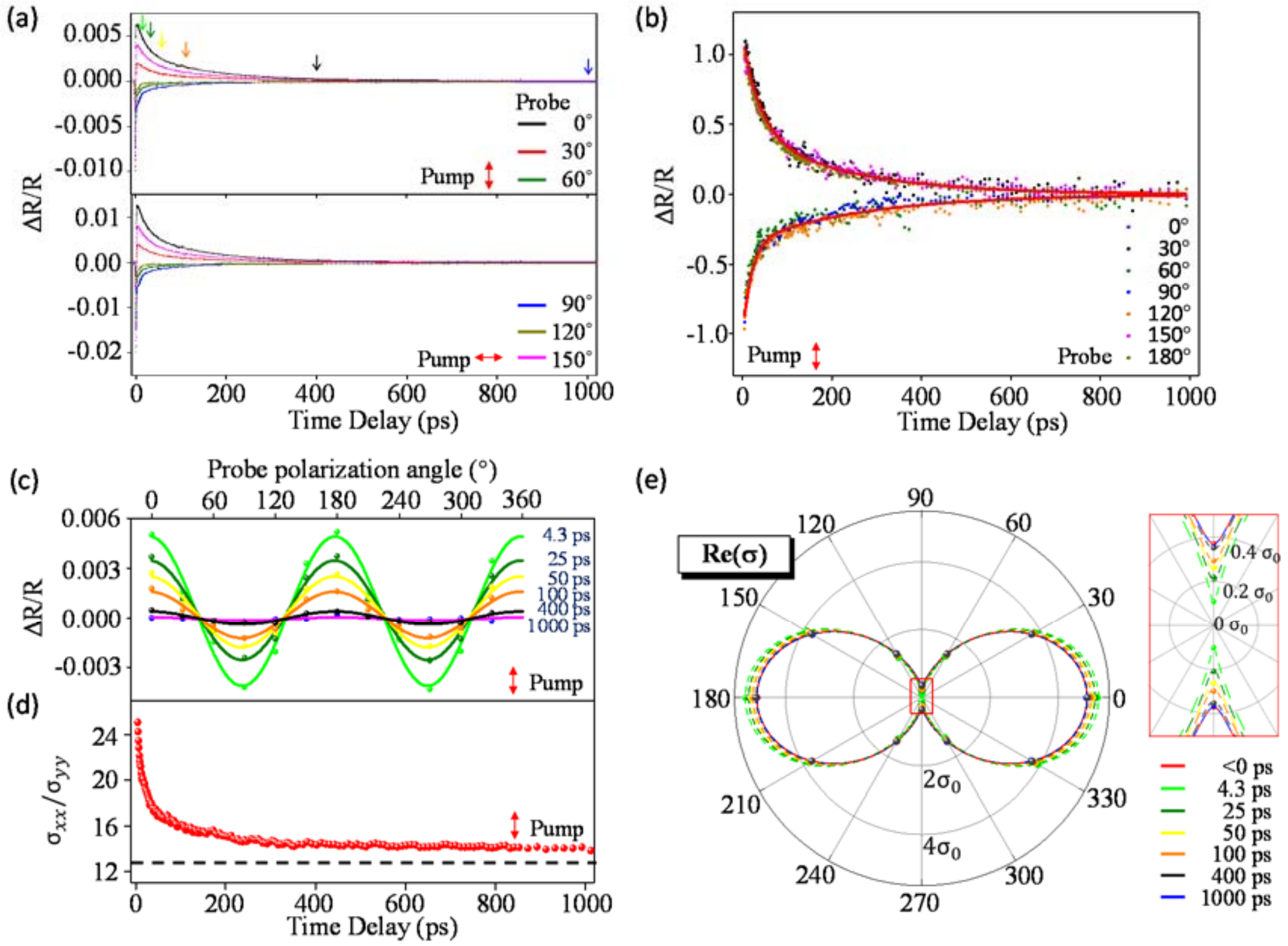

**Figure 3. Probe polarization dependence of transient reflection spectra,** (a) Probe polarization dependence of transient reflection spectrum with pump polarization fixed along 0˚ and 90˚. (b) Normalized transient reflection spectrum (normalized with $\Delta R/R|_{t=4.5\ \mathrm{ps}}$) with different probe polarization angle. (c) Transient reflection spectra with different probe polarization at fixed delays. The angle dependence is fit by function $\Delta R/R = a\cos^2\alpha + b\sin^2\alpha$. (d) Dynamic evolution of $\mathrm{Re}(\sigma_{xx})/\ \mathrm{Re}(\sigma_{yy})$ at different delays with pump polarized along 90˚(without subtracting constant background). The black dash line marks value of $\mathrm{Re}(\sigma_{xx})/\ \mathrm{Re}(\sigma_{yy})$ in a sample prior to pumping. (e) Dynamical evolution of Re (σ) ellipse at different delays. Δ Re (σ) is doubled for clarity. Inset: Enlarged plot of area marked by red rectangle.

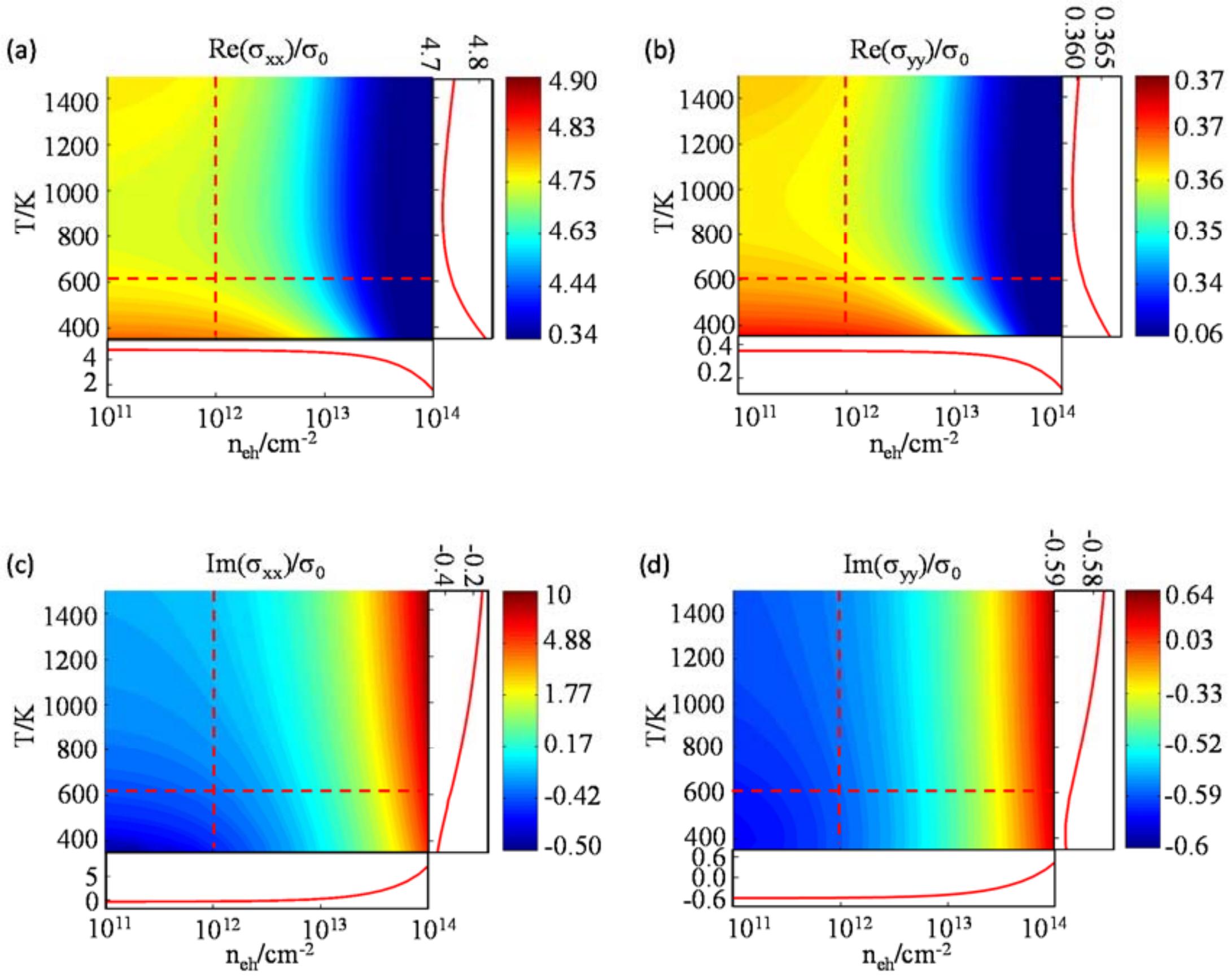

**<u>Figure 4</u>. Computed dynamical conductivity,** (a) $Re(\sigma_{xx}) / \sigma_0$ and (b) $Re(\sigma_{yy}) / \sigma_0$ as functions of photo-induced carrier density $n_{eh}$ and quasi-equilibrium temperature T. The inset on the right (bottom) shows a cut at fix $n_{eh} = 10^{12}/cm^2$ (T=600 K) as marked by red dash line. (c) $Im(\sigma_{xx}) / \sigma_0$ and (d) $Im(\sigma_{yy}) / \sigma_0$ as function of $n_{eh}$ and T. The inset on the right (bottom) shows a cut at fix $n_{eh} = 10^{12}/cm^2$ (T=600 K) as marked by red dash line.

# Supplementary Information

## Dynamical Evolution of Anisotropic Response in Black Phosphorus under Ultrafast Photo-excitation

Shaofeng Ge[1,2],Chaokai Li[1,2], Zhimin Zhang[1,2], Chenlong Zhang[1,2], Yudao Zhang[1,2], Jun Qiu[1,2], Qinsheng Wang[1,2], Junku Liu[3], Shuang Jia[1,2], Ji Feng[1,2,*], Dong Sun[1,2,*]

[1]International Center for Quantum Materials, School of Physics, Peking University, Beijing 100871, P. R. China
[2]Collaborative Innovation Center of Quantum Matter, Beijing 100871,P. R. China
[3]Qian Xuesen Laboratory of Space Technology, China Academy of Space Technology, Beijing 100094, P. R. China
[*]Email: jfeng11@pku.edu.cn (J. F.), sundong@pku.edu.cn (D.S.);

## *Supplementary Information*

Table of Contents:

## 1. Deduction of absorption coefficient A.

The reflectivity of the black phosphorus crystal is given in the literature[1]:

$$r = -\frac{\epsilon_0 c(\sqrt{\epsilon_2}-\sqrt{\epsilon_1})+\sigma_{xx}\cos^2(\alpha)+\sigma_{yy}\sin^2(\alpha)}{\epsilon_0 c(\sqrt{\epsilon_2}+\sqrt{\epsilon_1})+\sigma_{xx}\cos^2(\alpha)+\sigma_{yy}\sin^2(\alpha)} \quad (1)$$

where $\epsilon_0$ is the free-space permittivity, $\epsilon_1$ and $\epsilon_2$ are the relative permittivity of the air and $SiO_2$ in our sample geometry, c is the speed of light, and α is the light polarization angle respect to the x axis of the black phosphorus crystal.

The reflection and transmission coefficient R and T are given by:

$$R = |r|^2 \quad (2)$$

$$T = |1+r|^2\sqrt{\epsilon_2/\epsilon_1} \quad (3)$$

So we can get the absorption coefficient A:

$$A = 1 - T - R = \frac{4\epsilon_0 c\sqrt{\epsilon_1}(Re(\sigma_{xx})\cos^2(\alpha)+Re(\sigma_{yy})\sin^2(\alpha))}{[\epsilon_0 c(\sqrt{\epsilon_2}+\sqrt{\epsilon_1})+Re(\sigma_{xx})\cos^2(\alpha)+Re(\sigma_{yy})\sin^2(\alpha)]^2+[Im(\sigma_{xx})\cos^2(\alpha)+Im(\sigma_{yy})\sin^2(\alpha)]^2} \quad (4)$$

where $\sigma_0 = e^2/4\hbar$, $\epsilon_1 = 1, \epsilon_2 = 3.9$, $\epsilon_0 c = \frac{\sigma_0}{0.023} = 43.5\sigma_0$, Re(σ) and Im(σ) are real and imaginary part of σ respectively. For unexcited state of 25 nm sample with two band model simulation[1], $Re(\sigma_{xx})\sim 4.84\,\sigma_0$, $Im(\sigma_{xx})\sim -0.52\sigma_0$, $Re(\sigma) \ll \epsilon_0 c(\sqrt{\epsilon_2}+\sqrt{\epsilon_1})$ and $(Im(\sigma))^2 \ll [\epsilon_0 c(\sqrt{\epsilon_2}+\sqrt{\epsilon_1})]^2$, so the absorption can be simplified by the following:

$$A \approx \frac{4\sqrt{\epsilon_1}(Re(\sigma_{xx})\cos^2(\alpha)+Re(\sigma_{yy})\sin^2(\alpha))}{\epsilon_0 c(\sqrt{\epsilon_2}+\sqrt{\epsilon_1})^2} \quad (5)$$

which is proportional to the real part of complex conductivity.

## 2. Deduction of transient reflection ΔR/R.

Since $R = |r|^2$, assuming $\Delta\sigma \ll \sigma$, (the validity of this approximation can be identified from Fig. S6).

We can get

$$\frac{\Delta R}{R} = \frac{-4\epsilon_0 c\sqrt{\epsilon_1}\Delta Re(\sigma)\left[Im^2(\sigma)-(\epsilon_0 c\sqrt{\epsilon_2}+Re(\sigma))^2+\epsilon_0^2c^2\epsilon_1\right]}{\left[(\epsilon_0 c\sqrt{\epsilon_2}+\epsilon_0 c\sqrt{\epsilon_1}+Re(\sigma))^2+Im^2(\sigma)\right]\left[(\epsilon_0 c\sqrt{\epsilon_2}-\epsilon_0 c\sqrt{\epsilon_1}+Re(\sigma))^2+Im^2(\sigma)\right]}$$

$$+\frac{8\epsilon_0 c\sqrt{\epsilon_1}\Delta Im(\sigma)Im(\sigma)(\epsilon_0 c\sqrt{\epsilon_2}+Re(\sigma))}{\left[(\epsilon_0 c\sqrt{\epsilon_2}+\epsilon_0 c\sqrt{\epsilon_1}+Re(\sigma))^2+Im^2(\sigma)\right]\left[(\epsilon_0 c\sqrt{\epsilon_2}-\epsilon_0 c\sqrt{\epsilon_1}+Re(\sigma))^2+Im^2(\sigma)\right]} \quad (6)$$

Where $\sigma = \sigma_{xx}\cos^2(\alpha)+\sigma_{yy}\sin^2(\alpha)$. Similar to the last session, using $Im^2(\sigma) \ll \epsilon_0^2c^2$ and $Re(\sigma) \ll \epsilon_0 c$, the $\Delta R/R$ is given by:

$$\frac{\Delta R}{R} \approx \frac{4\sqrt{\epsilon_1}\Delta Re\sigma}{\epsilon_0 c(\epsilon_2-\epsilon_1)}+\frac{8\sqrt{\epsilon_1\epsilon_2}(\Delta Im\sigma)Im\sigma}{\epsilon_0^2c^2(\epsilon_2-\epsilon_1)^2} \quad (7)$$

If we plug in the values:

$$\frac{\Delta R}{R} \approx \frac{0.03}{\sigma_0}\Delta Re\sigma + \frac{0.001}{\sigma_0} * \frac{Im\sigma}{\sigma_0}\Delta Im\sigma \quad (8)$$

Taking Im $\sigma_x = -0.52\,\sigma_0$ and Im $\sigma_y = -0.60\,\sigma_0$ for native state of 25-nm thick BP sample from two band model simulation, the first term is usually dominates when $\Delta Re(\sigma)\sim$or $> \Delta Im(\sigma)$.

## 3. Bi-exponential fitting of pump power dependence of transient reflection

In Figure S1, we use bi-exponential function ΔR=A*(exp(-t/$\tau_1$)+β*exp(-t/$\tau_2$)) to fit the pump fluence dependence of transient reflection ΔR with probe polarization fixed at 90° (Fig.S1a) and 0° (Fig.S1b). The fitting starts from 4.3 ps delay to exclude the effect in non-equilibrium state closed to timezero. The decay time constants are tabulated in Table S1, both decay time constant $\tau_1$ and $\tau_2$ decrease as pump fluence increases.

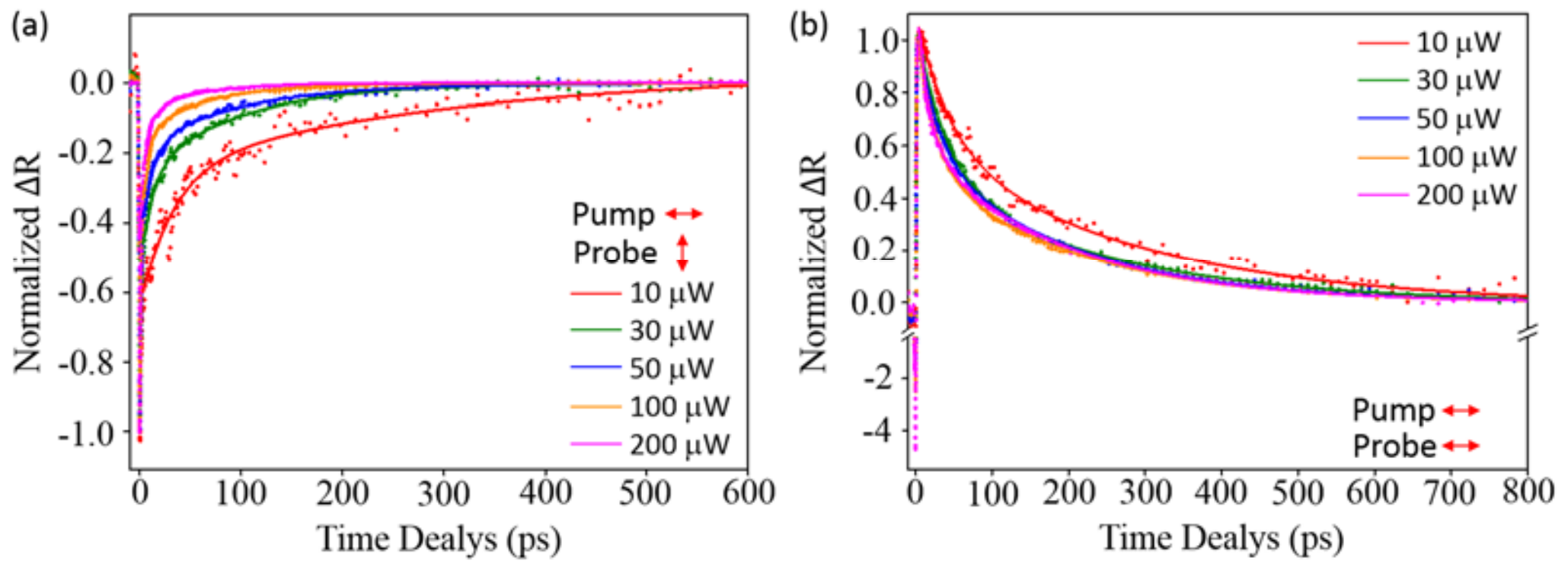

***Figure S1***: *Bi-exponential fitting of pump power dependent transient reflection data taken at 77K with probe polarization fixed along 90° (Fig.1 (a)) and 0° (Fig.1 (b)). The fitting parameter:* $\tau_1$,$\tau_2$ *and β for each pump fluence are shown in Table S1a and S1b for 90° and 0° respectively.*

| Pump power (μW) | 10 | 30 | 50 | 100 | 200 |
|---|---|---|---|---|---|
| $\tau_1$(ps) | 32.7 | 10.9 | 10.3 | 4.8 | 5.5 |
| $\tau_2$(ps) | 303.8 | 94.0 | 97.3 | 59.7 | 49.4 |
| β | 0.72 | 0.93 | 0.62 | 0.35 | 0.31 |

***Table S1(a)***:*Fitting parameters for different pump power with probe polarization* fixed along 90°.

| Pump power (μW) | 10 | 30 | 50 | 100 | 200 |
|---|---|---|---|---|---|
| $\tau_1$(ps) | 43.8 | 38.5 | 25.3 | 21.1 | 14.5 |
| $\tau_1$(ps) | 283.7 | 245.4 | 195.7 | 196.2 | 186.2 |
| β | 1.33 | 0.88 | 1.30 | 1.00 | 1.06 |

***Table S1(b)***: *Fitting parameters for different pump power with probe polarization* fixed along 0°.

## 4. Rate equation fitting of pump power dependence of transient reflection.

As shown in Fig. 2c and bi-exponential fitting of Fig. S1, the decay time constants of the transient reflection signal has strong dependence on the initial pump excitation intensity. To study this in detail, we model the process by decay kinetics in which we include three carrier recombination terms with different power order of carrier density by the following rate equation[2]:

$$\frac{dN_n(t)}{dt} = G(t) - AN_n(t) - BN_n^2(t) - CN_n^3(t) \qquad (9)$$

where $N_n(t)$ is the electron density, $A$ ($s^{-1}$), $B$ ($cm^3\ s^{-1}$), and $C$ ($cm^6\ s^{-1}$) accounts for different carrier density dependent recombination processes, for example, Schockley-Reed, electron-hole and Auger recombination coefficients[3], respectively. The generation rate, $G$ ($cm^{-3}\ s^{-1}$), is proportional to the absorbed laser excitation $q$ ($W/cm^2$) which is nonzero during the pump pulse duration.

The 100 μW excitation power converts to an absorbed fluence of $3.81\times 10^3$ μ $J/cm^2$ using 0.04% per layer absorption rate of BP at 800 nm[1]. Assuming that each absorbed photon generates one electron-hole pair, the resulting photoexcited carrier density is $3 \times 10^{14}\ cm^{-2}$. Within the range of applied pump excitation in our experiment, we find the pump induced transient reflection at timezero scales linearly with pump fluence as shown in Fig. 2 and Fig. S2b. According to this observation, we make the assumption that the absorption coefficient of the pump radiation remains independent of pump fluence, as expected for rapid relaxation of initially injected carrier and the bleaching of the sample remains linear with injected carrier density in the range of our experiment.

Here, we show that the entire set of data of the transient spectrum at different pump fluence can be fit by a simple model based on rate equation (9). For the multi-curve fitting, we numerically integrate the rate equation and then compute the rms deviation against the experimental data. Different weight has been given to the experimental data at different pump power and different time delay in order to make sure the data contribute equally to the rms. By varying the parameters to minimize the rms deviation, we get the Schockley-Reed, radiative and Auger recombination coefficients. This simple model of the decay dynamics provides a good fit to the complete set of experimental data, as indicated by the solid lines in Fig. 2c and Fig. S2a. Considering all sets of data, we obtain A=0.003546 $s^{-1}$, B=$2.74\times 10^{-21}\ cm^3 s^{-1}$, C= $4.098\times 10^{-43}\ cm^6 s^{-1}$. We note the fitting here only provides suggestions of relative proportion of recombination processes that have different power orders of carrier densities, the absolute value of the fitting parameter may deviate from reality significantly due to simplification of the model and other effects may come into play.

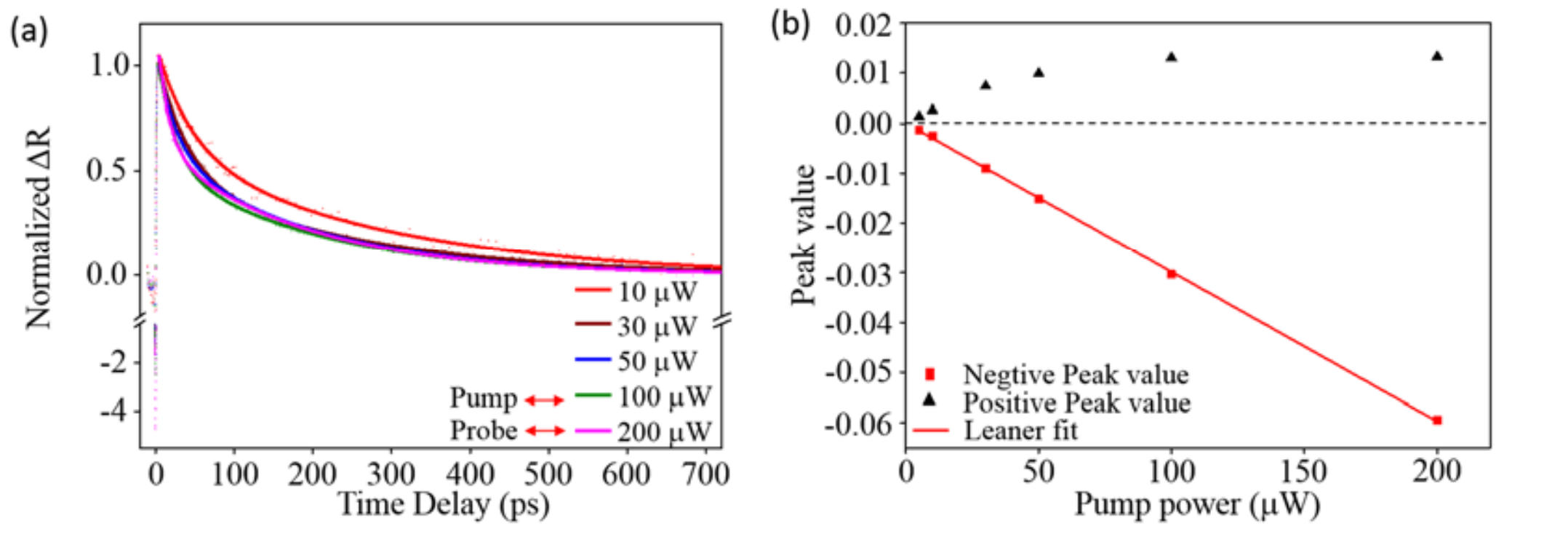

***Figure S2***: *(a) Normalized transient reflection at different pump excitation power taken at 77K when both pump and probe polarization is fixed along 0°. (b) Peak value of the transient reflection spectrum vs. pump power. The red line is the leaner fit of the experiment data.*

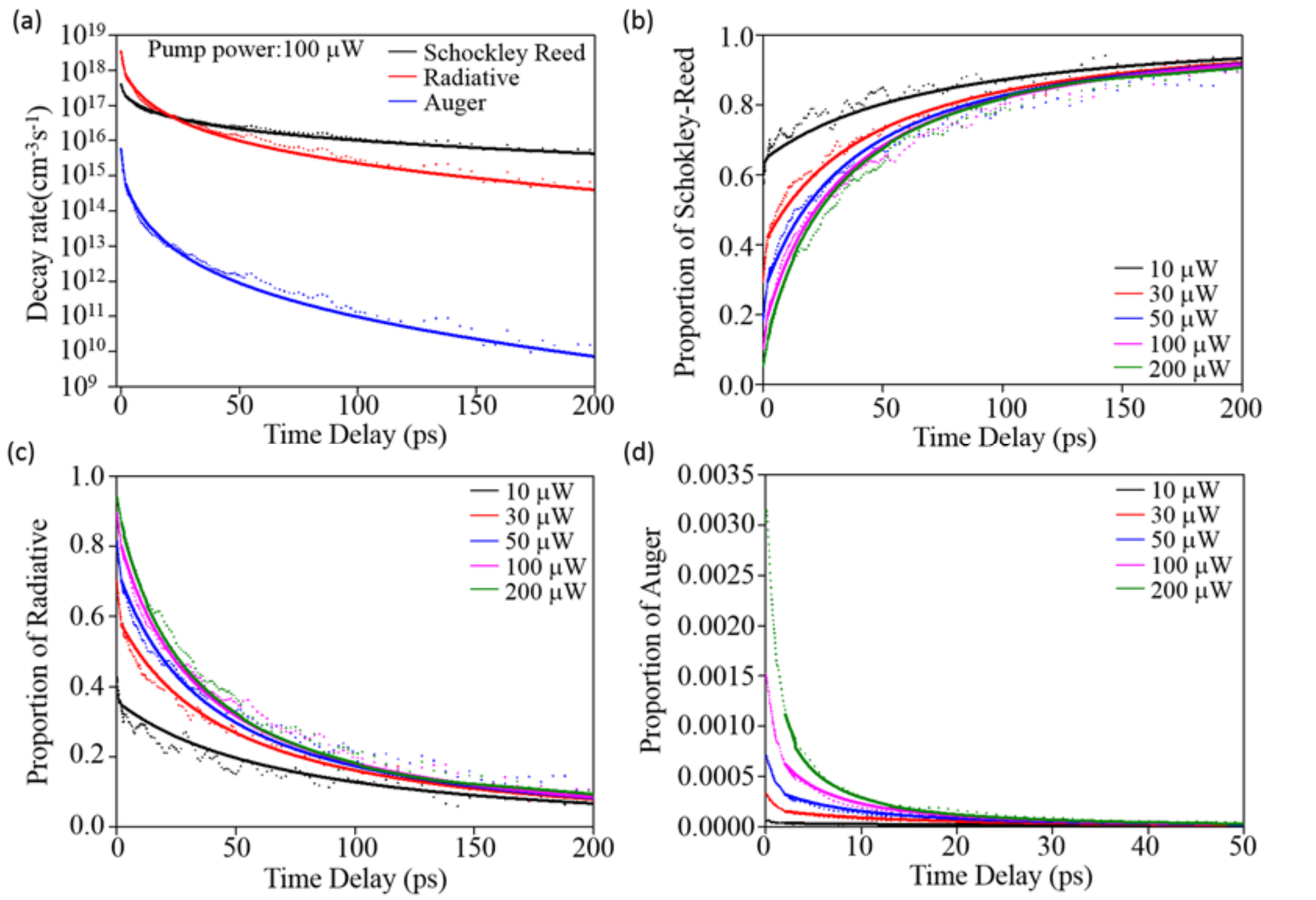

***Figure S3***: *(a) Carrier decay rate of Schockley-Reed, radiative, Auger recombination process with pump power of 100 μW. (b) The proportion of Schockley-Reed process with different pump influences. (c) The proportion of radiative process with different pump influence. (d) The proportion of Auger process with different pump influence.*

Using the recombination coefficients obtained by fitting the experimental results, we can plot the dynamical evolution of relative contribution of different recombination processes. As shown in Figure S3a. The Schockley-Reed and the electron-hole recombination are the dominate decay channels during the entire decay, both processes are 2 orders of magnitude larger than the Auger process. All three processes slow down as the carrier density decreases, while the decay rate of Auger process decays faster as it is more sensitive to the carrier density than the others. The proportion of the Schockley-Reed, electron-hole and Auger recombination processes are shown in Fig. S3b,c,d respectively. We can see as the pump fluence increases, the probability of the radiative and the Auger process increases but the proportion of the Schockley-Reed process decreases. The Auger process is only obvious at high carrier densities around time zero.

## 5. Pump-probe polarization and temperature dependence of transient reflection background

As shown in the Fig. 1e, the transient reflection signal has a long-lived "background" after even nanosecond delay. We tentatively attribute this background signal to the elevated temperature of the sample compared to the environment, and thus the slow decay of the background to dissipation of heat to the environment (substrate). Figure S4 shows the background signals as a function of probe and pump polarization angle taken at room temperature. For the probe polarization dependent measurement (with pump polarization fixed along 90°), the results can be fitted by $asin^2(\alpha - \varphi) + bcos^2(\alpha - \varphi)$, with $\varphi$ = -5.3° (red curve), which is consistent with $\varphi$ fit from pump polarization dependent measurement shown in Fig. 3 of main text. For the pump polarization dependent measurement (with probe polarization fixed along 90°), the results show negligible no dependence on the pump polarization angle.

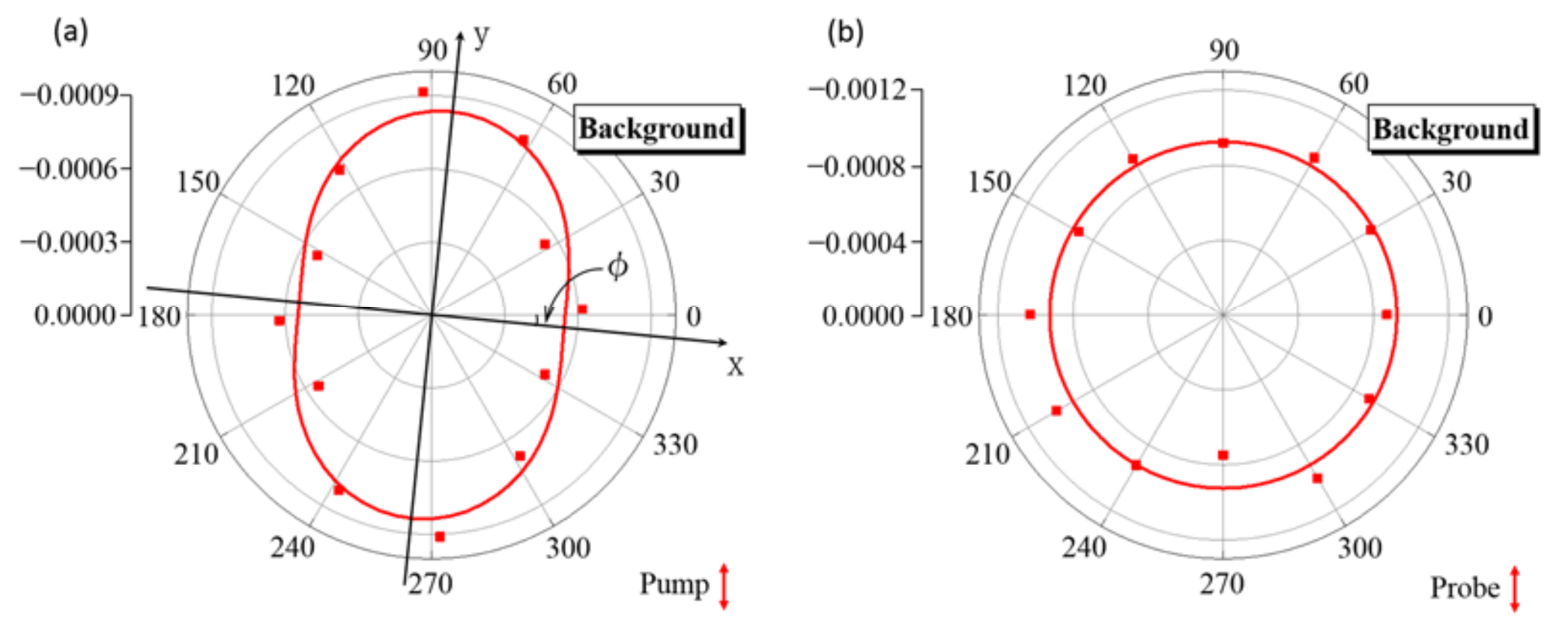

***Figure S4**: (a) Probe polarization dependence of background signal with pump polarization fixed along 90°. (b) Pump polarization dependence of background signal with probe polarization fixed along 90°.*

Figure S5 shows the transient reflection background dependence on the temperature with probe polarization along 0° (Fig. S5a) and 90° (Fig. S5b). In either case, the pump polarization is fixed along 0°. The result indicates that the background decreases as the temperature increases in both configurations. However, when probe polarization is along 0°, the background experiences a sign switch when the temperature increases from 77 K to 290 K. The observed temperature dependence supports that the background is due to the elevation of lattice temperature. The sample remains at slightly elevated lattice temperature due to the heating of pump pulse over period, and it takes well over a nanosecond to cool down to the environment temperature by heat transfer to the substrate.

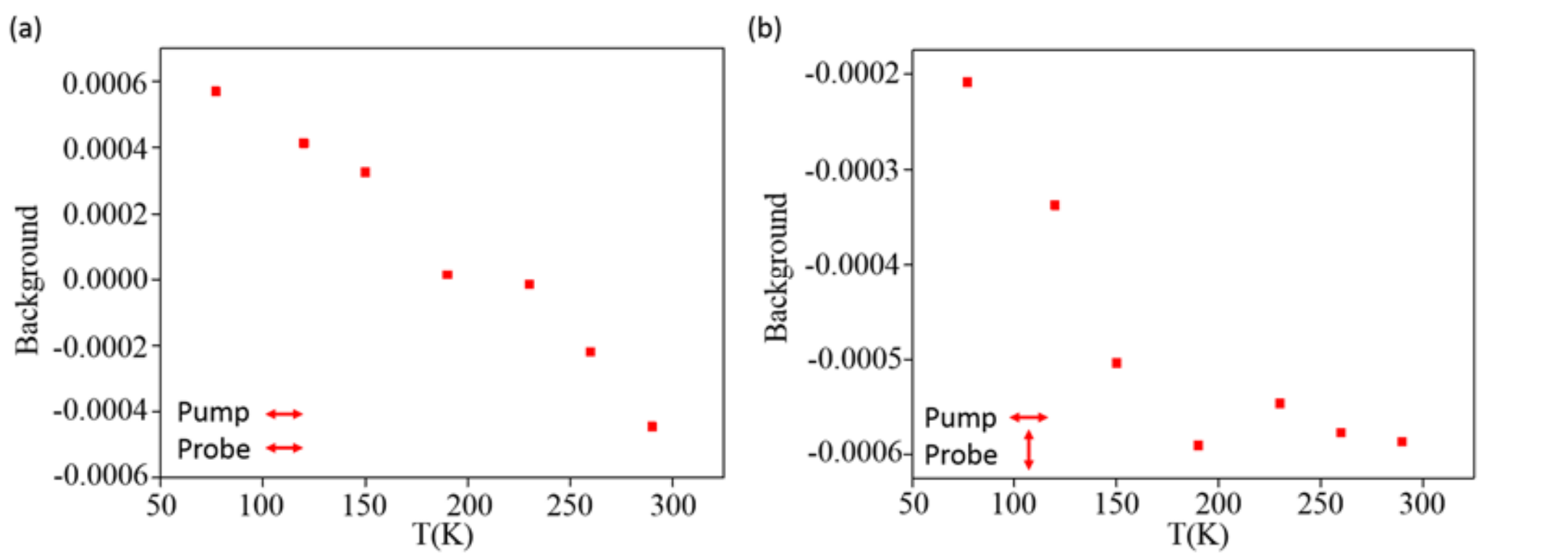

***Figure S5***: *(a) Temperature dependence of background with both pump and probe polarization fixed along 0°. (b) Temperature dependence of background with pump and probe polarization fixed along 0° and 90°.*

## 6. Relative contribution of real and imaginary parts of conductivity changes

To examine in which regime the Taylor expansion used to derive Eq. (3) is valid, we define the relative error of the expansion as

$$\text{error} = \frac{(\Delta R/R)'}{\Delta R/R} - 1 \quad (10)$$

where $\Delta R/R$ on the denominator and $(\Delta R/R)'$ on the numerator are, respectively, the values before and after Taylor expansion, as given in Eq. (3). As can be seen from Figs. S6 a and b, the error of Eq. (3) in the main text is less than 15% in the range of $10^{11}\text{cm}^{-2} < n_{eh} < 10^{13}\text{cm}^{-2}$ and $400\text{K} < T < 1500K$. Furthermore, Figs. S6c and d show that the ratio of the second term to the first term of eq.(7) is less than 10% in the same regime. This indicates that in this range the change of the real part of conductivity dominates the contribution to ΔR/R, validating the approximation used in Eq.(3) of the main text.

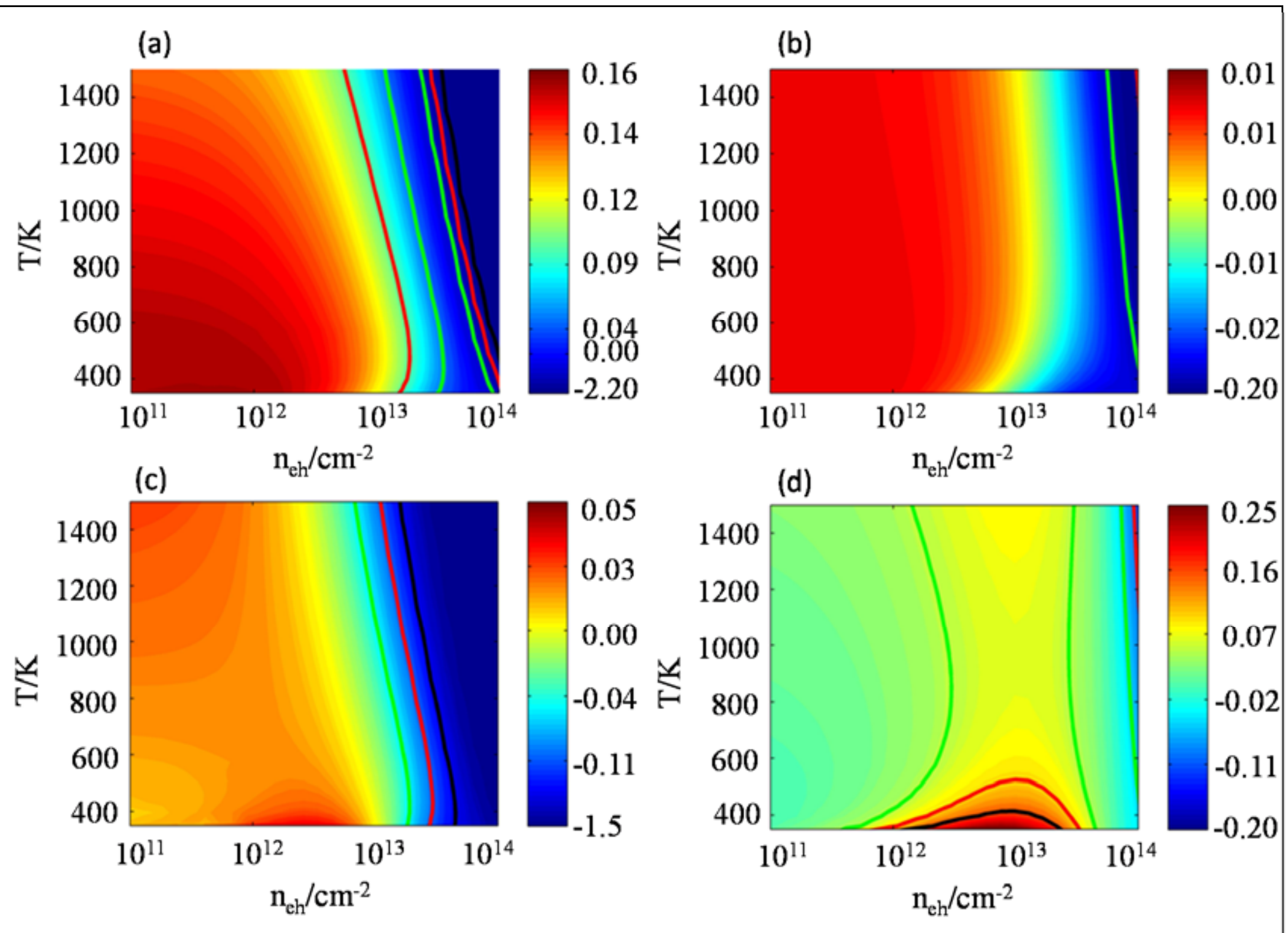

***Figure S6***: *Relative error (as defined by Eq. (10)) of in the Taylor expansion for obtaining Eq.(3) in the main text for x-polarization (a) and y-polarization (b). The ratio or the second term to the first term of Eq.(3) of the main text for x-polarization (c) and y-polarization (d). The black, red, and green curves mark the contours of ±16%, ±10% ,and ±5%, respectively.*

## 7. Supplementary Reference